\documentclass[letterpaper]{article} % DO NOT CHANGE THIS
\usepackage{aaai2026}  % DO NOT CHANGE THIS
\usepackage{times}  % DO NOT CHANGE THIS
\usepackage{helvet}  % DO NOT CHANGE THIS
\usepackage{courier}  % DO NOT CHANGE THIS
\usepackage[hyphens]{url}  % DO NOT CHANGE THIS
\usepackage{graphicx} % DO NOT CHANGE THIS
\usepackage{natbib}  % DO NOT CHANGE THIS AND DO NOT ADD ANY OPTIONS TO IT
\usepackage{caption} % DO NOT CHANGE THIS AND DO NOT ADD ANY OPTIONS TO IT
\usepackage{algorithm}
\usepackage{algorithmic}

\usepackage{multirow}
\usepackage{subfigure}
\usepackage{arydshln}

\usepackage{newfloat}
\usepackage{listings}
\DeclareCaptionStyle{ruled}{labelfont=normalfont,labelsep=colon,strut=off} % DO NOT CHANGE THIS
\floatstyle{ruled}
\newfloat{listing}{tb}{lst}{}
\floatname{listing}{Listing}
\title{Towards Better Code Understanding in Decoder-Only Models with Contrastive Learning}
\author{
    Jiayi Lin\textsuperscript{\rm 1,2}, Yanlin Wang\textsuperscript{\rm 3}\thanks{Corresponding author.}, Yibiao Yang\textsuperscript{\rm 4}, Lei Zhang\textsuperscript{\rm 1}, Yutao Xie\textsuperscript{\rm 1}
}
\affiliations{
    \textsuperscript{\rm 1}International Digital Economy Academy, Shenzhen, China\\
    \textsuperscript{\rm 2}Tencent, Shenzhen, China\\
    \textsuperscript{\rm 3}School of Software Engineering, Sun Yat-sen University, Zhuhai, China\\
    \textsuperscript{\rm 4}State Key Laboratory for Novel Software Technology, Nanjing University, Nanjing, China\\
    jiayilin1024@gmail.com, yanlin-wang@outlook.com, yangyibiao@nju.edu.cn, \{leizhang,yutaoxie\}@idea.edu.cn
}

\begin{document}

\maketitle

\begin{abstract}
Recent advances in large-scale code generation models have led to remarkable progress in producing high-quality code. These models are trained in a self-supervised manner on extensive unlabeled code corpora using a decoder-only architecture. However, despite their generative strength, decoder-only models often exhibit limited performance on code understanding tasks such as code search and clone detection, primarily due to their generation-oriented training objectives. While training large encoder-only models from scratch on massive code datasets can improve understanding ability but remains computationally expensive and time-consuming.
In this paper, we explore a more efficient alternative by transferring knowledge from pre-trained decoder-only code generation models to code understanding tasks. We investigate how decoder-only architectures can be effectively adapted to learn discriminative and semantically meaningful code representations. To this end, we propose CL4D, a contrastive learning framework tailored to strengthen the representation capabilities of decoder-only models.
Extensive experiments on multiple benchmark datasets demonstrate that CL4D achieves competitive or superior performance compared to existing methods on representative code understanding tasks, including code search and clone detection. Further analysis reveals that CL4D substantially improves the semantic alignment of code representations by reducing the distance between semantically similar code snippets. These findings highlight the feasibility of leveraging decoder-only models as a unified backbone for both code generation and understanding.
\end{abstract}

\begin{links}
    \link{Code}{https://github.com/JiayiLin1024/CL4D}
\end{links}

\section{Introduction}
Recent years have witnessed a surge of large-scale pre-trained code generation models. 
These models are trained in a self-supervised manner on vast code corpora using a decoder-only architecture and have demonstrated remarkable success~\cite{DBLP:conf/iclr/NijkampPHTWZSX23, DBLP:journals/corr/abs-2401-14196, DBLP:journals/corr/abs-2402-19173}. 
They have been widely adopted in developer-assistance tools, such as GitHub Copilot\footnote{https://github.com/features/copilot} and Cursor\footnote{https://cursor.com/}, to support everyday programming tasks. 

\begin{figure}[t]
  \includegraphics[width=\columnwidth]{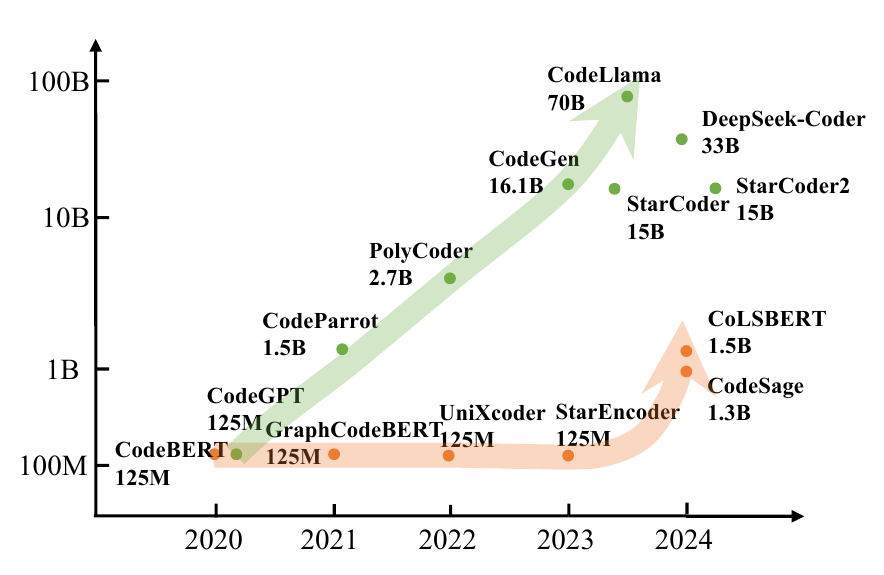}
  \caption{Timeline of code-large models. 
  For visual clarity, only the largest size of representative models is plotted. 
  The orange arrow indicate the development trends of encoder-only code understanding models, 
  while the green arrow represent the trends of decoder-only code generation models.}
  \vspace{-1.5em}
  \label{fig:model_size}
\end{figure}

Despite their outstanding generative performance, decoder-only models often struggle with code understanding tasks, including code search and code clone detection. This limitation arises from their auto-regressive training objectives, which primarily emphasize next-token prediction rather than semantic comprehension. 
Previous studies have shown that stronger performance on code understanding tasks can be achieved by pre-training encoder-only models from scratch on large-scale code datasets~\cite{DBLP:journals/corr/abs-2203-15556, DBLP:journals/corr/abs-2402-12813}. 
However, such training is computationally expensive and time-consuming, making it impractical for widespread adoption.  
As illustrated in Figure~\ref{fig:model_size}, the landscape of code models has evolved rapidly: from 1.1B SantaCoder~\cite{DBLP:journals/corr/abs-2301-03988} and the 15B StarCoder~\cite{DBLP:journals/corr/abs-2305-06161} to the 70B CodeLlama~\cite{DBLP:journals/corr/abs-2308-12950}. 
These models are significantly larger and more data-rich than the current leading code understanding models, 
such as the 1.3B CodeSage~\cite{zhang2024codesage} and the 1.5B CoLSBERT~\cite{DBLP:journals/corr/abs-2402-12813}. 
Nevertheless, how to effectively leverage these powerful pre-trained decoder-only code generation models for code understanding tasks remains an open and unexplored problem. 

In this paper, 
we investigate the adaption of pre-trained decoder-only code generation models to code understanding tasks. 
We explore two methods for extracting code representations from the decoder-only Transformer: 
1) using the encoding of the last token, 
and 2) using the average encodings of all tokens. With these code representations, we introduce CL4D, 
a contrastive learning framework that continues pre-training the code generation models to enhance their representational capabilities.  
Follow Gao et al.~\cite{DBLP:conf/rep4nlp/GaoZHC21}, we employ hard negatives, semantically distinct code snippets that are close to the query in representation space, to improve training efficiency. 

To ensure rigorous evaluation, 
we compare pre-trained decoder-only and encoder-only models of comparable size on two representative understanding tasks: code search and clone detection. 
Experimental results indicate that our proposed CL4D effectively adapts decoder-only code generation models for code understanding tasks. 
After continued pre-training on a small code corpus, 
these models outperform pre-trained encoder-only models on most tasks. 
Furthermore, we observe that larger decoder-only models further improve code understanding performance.

In summary, the contributions of this paper are as follows:
\begin{itemize}
  \item We investigate how pre-trained decoder-only code generation models can be effectively leveraged for code understanding tasks, bridging the gap between generation-oriented and understanding-oriented applications.
  \item We propose CL4D, a contrastive learning–based continued pre-training approach that enhances the representation ability of decoder-only models, enabling competitive or superior performance to encoder-only models on code understanding benchmarks.
  \item Our findings reveal that larger decoder-only models benefit code understanding tasks and demonstrate the potential for unifying code understanding and code generation within a single model architecture.
\end{itemize}

\begin{figure*}[t]
  \centering
  \includegraphics[width=\textwidth]{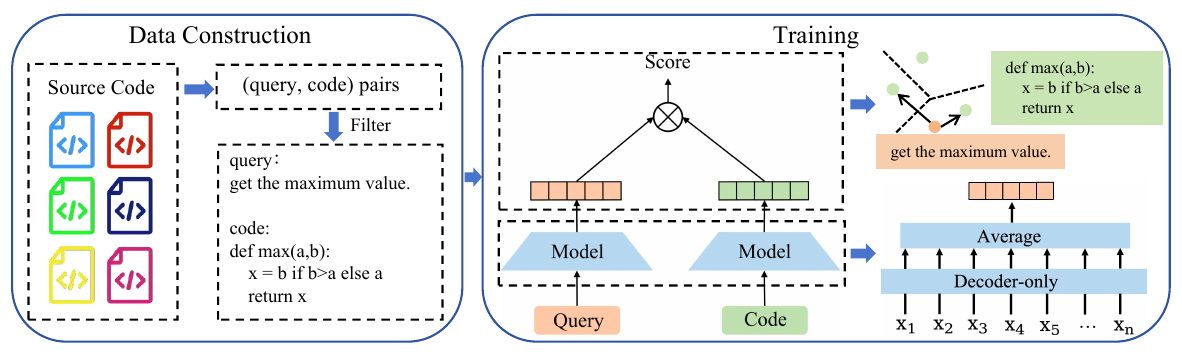}
  \caption{Overview of CL4D.}
  \label{fig:overview}
\end{figure*}

\section{Background and Motivation}
In recent years, large-scale language models with a decoder-only architecture have achieved remarkable success in the field of code generation. Models such as Codex~\cite{DBLP:journals/corr/abs-2107-03374}, CodeGen~\cite{DBLP:conf/iclr/NijkampPHTWZSX23}, StarCoder~\cite{DBLP:journals/corr/abs-2305-06161}, and Code Llama~\cite{DBLP:journals/corr/abs-2308-12950} have demonstrated impressive capabilities in generating syntactically correct and semantically meaningful code. These models are typically pre-trained on massive corpora of natural language and programming languages, allowing them to acquire extensive programming knowledge and multi-language fluency. With the emergence of instruction-tuned variants like WaveCoder~\cite{DBLP:conf/acl/YuZSHXZHY24} and CodeGemma~\cite{DBLP:journals/corr/abs-2406-11409}, decoder-only models continue to push the boundaries of generative performance.

Despite their superiority in generative tasks, our empirical observations reveal a surprising performance gap when these models are applied directly to code understanding tasks such as code search and code clone detection. As shown in Table~\ref{CL4D_res}, encoder-only models—such as GraphCodeBERT~\cite{DBLP:conf/iclr/GuoRLFT0ZDSFTDC21}, UniXcoder~\cite{DBLP:conf/acl/GuoLDW0022} and CodeSage~\cite{zhang2024codesage}—which are explicitly designed for understanding source code, consistently outperform their decoder-only counterparts on benchmark code understanding datasets.

This performance gap persists even though decoder-only models possess several natural advantages. First, they have been exposed to significantly larger and more diverse code corpora during pre-training, often spanning multiple programming languages and domains. Second, these models are frequently trained with objectives aligned with next-token prediction, which implicitly teaches structural and semantic patterns in code. However, their lack of bi-directional attention mechanisms and their generative training objective may limit their capacity for fine-grained code comprehension.

One obvious solution is to pre-train a new encoder-only model at the same scale as the latest decoder-only models. However, the computational and financial costs of such an endeavor are prohibitively high, particularly when targeting billion-parameter models trained over trillions of tokens. This raises a critical research question: Can we enhance the code comprehension capabilities of existing decoder-only models without incurring the massive cost of training new models from scratch?

To this end, we propose a contrastive learning framework aimed at improving the code understanding performance of decoder-only models. Our approach leverages the inherent strengths of pre-trained decoder-only models while avoiding the prohibitive cost of full model retraining.

\section{Methodology}
\label{sec:method}
In this section, 
we explore obtaining encoding from decoder-only pre-trained code generation models to address code understanding tasks.
The overview of CL4D is shown in Figure~\ref{fig:overview}. 
We first introduce the data construction method and then introduce the training of CL4D, including the model architecture and contrastive learning.

\subsection{Data Construction}
\label{sec:data_construction}
We train decoder-only Transformer models on six programming languages from The Stack dataset~\cite{DBLP:journals/corr/abs-2211-15533}: Python, Java, Go, PHP, JavaScript, and Ruby. 
To enhance the ability of model to differentiate between natural language and programming language, 
we use Tree-Sitter\footnote{https://tree-sitter.github.io/tree-sitter/} to extract bimodal data from The Stack dataset for training, i.e., $(query, code)$ pairs. 
The query is the first line of the function docstring, which approximates the query of user. 
Additionally, 
we apply the filtering rules to improve the quality of the training corpus following CSN~\cite{DBLP:journals/corr/abs-1909-09436}.
Given that pre-trained code generation models have already learned a vast amount of code knowledge, 
we construct millions of data samples for continued pre-training.

\subsection{Model Architecture}
We investigated two methods for obtaining code representation from decoder-only models: 
1) using the embedding of the last token, 
and 2) averaging the embeddings of all tokens. 
Encoder-only models like BERT typically use the embedding of the first token $[CLS]$ in the last hidden layer as the feature representation of the input code snippet, 
utilizing a bidirectional attention mechanism. 
By contrast, 
decoder-only models use a unidirectional attention mechanism, 
where only subsequent tokens can refer to previous information. 
Consequently, only the last token aggregates the information of the entire sample. 
In the first method, we use the embedding of the last token in the last hidden layer as the feature representation of the input code snippet. 
In the second method, we calculate the average of the embeddings of all input tokens in the last hidden layer as the feature representation.

Following previous work~\cite{DBLP:conf/iclr/GuoRLFT0ZDSFTDC21, zhang2024codesage}, 
we adopt a dual-encoder architecture to continue pre-training the code encoding model. 
As depicted in Figure~\ref{fig:overview}, 
this architecture consists of two Transformer decoder modules with shared weights to encode the input code snippet. 
In our experiments, we initialize the Transformer decoder module with the pre-trained code generation models.

\subsection{Contrastive Learning}
The traditional left-to-right training method for predicting the next token limits the representation capability of code generation models. 
To fully leverage the extensive code knowledge in the existing open-source code generation models and to avoid the high costs of retraining code understanding models, 
we employ contrastive learning to further fine-tune these models, 
thereby improving their representation abilities.

Contrastive learning is the primary method for enhancing the representation ability of models~\cite{DBLP:conf/emnlp/ZhangLH23}.
Here, we employ contrastive learning tasks to train pre-trained code generation models, 
addressing the limited representation ability of decoder-only Transformer due to the unidirectional attention mechanism.

Specifically, 
we randomly sample n samples from the training corpus to form a batch of data $B={(q_1,c_1),...,(q_n, c_n)}$.
For a $(q_i, c_i)$ pair,
we take $c_i$ as the positive sample $c^+$ for $q_i$,
and the other $n-1$ codes in the batch, excluding $c_i$, as negative samples for $q_i$.
Additionally, 
following Gao et al.~\cite{DBLP:conf/rep4nlp/GaoZHC21}, we construct hard negative samples $c^h$ for $q_i$.
Specifically, during the data processing stage, 
we utilize UniXcoder to select a code snippet from the entire codebase for each query in the training set, i.e., the code snippet that is close to the query in the representation space but have different semantic from the query. 
The loss function can be formulated as:
\begin{equation} 
\resizebox{.55\hsize}{!}{
$\mathcal{L}=-\log \frac{exp\left(s\left(q, c^+\right) / \tau \right)}{\sum_{i=1}^n exp\left(s\left(q, c_i\right) / \tau \right) + exp\left(s\left(q, c^h\right) / \tau \right)},$
}
\end{equation}
where $\tau$ is the temperature hyper-parameter, which we set to 0.05. 
$s(q,c)$ denotes the correlation score between the two, calculated using cosine similarity. 
The training objective is to minimize the distance between related pairs of samples in the representation space while maximizing the distance between unrelated pairs. 
Each batch is randomly sampled from the entire corpus, 
allowing for mixed programming languages within a batch, 
which helps the pre-trained model learn a unified semantic space for codes in different programming languages.

\section{Experimental Setup}

In this section, we conduct experiments to address the following research questions:
\begin{itemize}
  \item RQ1: Can our proposed CL4D boost the performance of the decoder-only code generation models for the code understanding tasks?
  \item RQ2: Which method is more suitable for a decoder-only model to extract code representation?
  \item RQ3: To what extent does contrastive learning enhance the code representations of decoder-only models?
  \item RQ4: How does CL4D perform in the zero-shot setting?
  \item RQ5: Why does our approach work?
\end{itemize}

\subsection{Understanding Tasks}
We evaluate the code understanding capabilities of decoder-only Transformer on two primary tasks:
\begin{itemize}
  \item \textbf{\textit{Code Search}}: 
  This task aims to find the most semantically similar code snippet in a candidate codebase based on a given natural language query. 
  We perform experiments on two datasets: CodeSearchNet (CSN)~\cite{DBLP:journals/corr/abs-1909-09436} and CoSQA~\cite{DBLP:conf/acl/HuangTSG0J0D20}. 
  CSN is a multi-language code corpus collected from GitHub, including Python, Java, Go, PHP, JavaScript, and Ruby. 
  CoSQA contains real user queries from Microsoft Bing search logs, with code snippets from CSN. 
  We use Mean Reciprocal Rank (MRR) as the evaluation metric, which is widely adopted in previous studies~\cite{DBLP:conf/acl/GuoLDW0022, zhang2024codesage}.
  \item \textbf{\textit{Clone Detection}}: 
  This task aims to identify the most semantically similar code snippet in a candidate codebase based on a given code snippet as query. 
  We conduct experiments on the POJ-104 dataset~\cite{DBLP:conf/aaai/MouLZWJ16}, designed to measure semantic similarity between code snippets. 
  Mean Average Precision (MAP) is the evaluation metric for this task.
\end{itemize}

\subsection{Baselines}
We compare the code understanding capabilities of encoder-only models and decoder-only models of comparable sizes, and evaluate the extend to which the proposed CL4D method enhances the performance of decoder-only models.

For encoder-only models, we select four state-of-the-art pre-trained models:
\begin{itemize}
  \item \textbf{\textit{CodeBERT}}~\cite{DBLP:conf/emnlp/FengGTDFGS0LJZ20} (125M) is trained with masked language modeling and replaced token detection on six programming languages in CodeSearchNet.
  \item \textbf{\textit{GraphCodeBERT}}~\cite{DBLP:conf/iclr/GuoRLFT0ZDSFTDC21} (125M) is trained with additional code structure understanding tasks, including edge prediction and node alignment.
  \item \textbf{\textit{UniXcoder}}~\cite{DBLP:conf/acl/GuoLDW0022} (125M) is trained with multiple language modeling and contrastive learning tasks. It is an encoder-decoder architecture model, where only the encoder is used for code understanding tasks via a special token $<encoder\_only>$.
  \item \textbf{\textit{CodeSage}}~\cite{zhang2024codesage} (1.3B) is trained with masked language modeling and contrastive learning on the extensive code corpus from The Stack dataset.
\end{itemize}

For decoder-only models, we select four state-of-the-art pre-trained models:
\begin{itemize}
  \item \textbf{\textit{CodeGPT}}~\cite{DBLP:conf/nips/LuGRHSBCDJTLZSZ21} (125M) is trained on CodeSearchNet with the same structure and training objectives as GPT-2~\cite{radford2019language}.
  \item \textbf{\textit{CodeGen}}~\cite{DBLP:conf/iclr/NijkampPHTWZSX23} (350M) is a model for program analysis trained on extensive natural language and programming language data. We choose the 350M version to facilitate comparison with similarly sized encoder-only models.
  \item \textbf{\textit{SantaCoder}}~\cite{DBLP:journals/corr/abs-2301-03988} (1.1B) is trained with multi-query attention and fill-in-the-middle on Java, JavaScript, and Python subsets of The Stack, incorporating near-deduplication and comments-to-code filtering.
  \item \textbf{\textit{phi-1}}~\cite{DBLP:journals/corr/abs-2306-11644} (1.3B) is a code generation model trained using textbook data sourced from the internet and textbooks synthesized by GPT-3.5.
  \item \textbf{\textit{DeepSeek-Coder}}~\cite{DBLP:journals/corr/abs-2401-14196} (1.3B) is a model trained on a high-quality project-level code corpus.
\end{itemize}

\subsection{Training details}
During the pre-training phase, we utilize the tokenizer from each decoder-only model directly for tokenization.
These models are trained using the AdamW optimizer, with a learning rate set to 2e-5. 
We train the decoder-only models for 2 epochs on the dataset, employing a batch size of 64. 
The experiments are conducted on 8 GPUs (80G A100). 
The longest pre-training duration is observed for phi-1, which takes approximately 3 days. 
In the fine-tuning phase, we adhere to the experimental settings of UniXcoder to minimize the impact of hyper-parameters on the fine-tuning results.

\section{Evaluation}
\subsection{RQ1: Overall performance comparison}
We evaluate the effectiveness of CL4D on two representative code understanding tasks: code search and clone detection. 
The experiment results are shown in Table~\ref{fine-tuned_result}, 
we can find that CL4D enables the decoder-only model to achieve state-of-the-art performance. 
Specifically, our method improves the performance of decoder-only models by approximately 2\% compared to encoder-only models of the same size. 
Additionally, as the size of the decoder-only model increases, its performance on downstream code understanding tasks shows a corresponding improvement. 
Overall, our approach effectively enhances the code understanding capabilities of decoder-only models.

\begin{table*}
  \centering
  \resizebox{0.85\textwidth}{!}{
  \begin{tabular}{llccc}
    \hline
    \multirow{2}{*}{}            & \multirow{2}{*}{Method}                                    & \multicolumn{2}{c}{Code Search}       &  Clone Detection \\
    \cline{3-5}
                                 &                                                                     & CSN            &  CoSQA      &  POJ-104 \\
    \hline
    \multirow{4}*{Encoder-only}  & CodeBERT~\cite{DBLP:conf/emnlp/FengGTDFGS0LJZ20} (125M)             & 70.18                    &  65.7                &  83.79               \\
                                 & GraphCodeBERT~\cite{DBLP:conf/iclr/GuoRLFT0ZDSFTDC21} (125M)        & 72.08                    &  68.4                &  85.50               \\
                                 & UniXcoder~\cite{DBLP:conf/acl/GuoLDW0022} (125M)                    & 74.40                    &  70.1                &  89.56               \\
                                 & CodeSage~\cite{zhang2024codesage} (1.3B)                            & 75.80                    &  68.0                  &  87.70               \\
    \hline
    \multirow{4}*{Decoder-only}  & CodeGPT~\cite{DBLP:conf/nips/LuGRHSBCDJTLZSZ21} (125M) + CL4D       & 70.20                    &  69.0                  & 87.96           \\
                                 & CodeGen~\cite{DBLP:conf/iclr/NijkampPHTWZSX23} (350M) + CL4D        & 73.30                    &  71.5                &  89.68           \\
                                 & SantaCoder~\cite{DBLP:journals/corr/abs-2301-03988} (1.1B) + CL4D   & 74.98                    &  72.2                &  83.98           \\
                                 & phi-1~\cite{DBLP:journals/corr/abs-2306-11644} (1.3B) + CL4D        & 75.18                    &  \textbf{72.8}                &  \textbf{92.72}           \\
                                 & DeepSeek-Coder~\cite{DBLP:journals/corr/abs-2401-14196} (1.3B) + CL4D                                &  \textbf{77.57}                   &  71.9                              &  89.71    \\
    \hline
  \end{tabular}
  }
  \caption{Comparison of the performance of encoder-only models and decoder-only models on code understanding tasks.}
  \label{fine-tuned_result}
\end{table*}

\subsection{RQ2: Code representation extraction method study for decoder-only models}

\begin{table*}[ht]
  \centering
  \resizebox{0.8\textwidth}{!}{
  \begin{tabular}{lcccccccc}
    \hline
    \multirow{2}{*}{Type}  & \multicolumn{7}{c}{CSN}            & \multirow{2}{*}{CoSQA} \\
    \cline{2-8}           
                              & Python         & Java        & Go        & Php     & JavaScript     & Ruby    & Avg     &  \\
    \hline
    Left padding + Avg        & 68.0                    &  67.7                &  89.1               &   62.9               &   59.5                      &      62.2            &   68.23              &     66.0           \\
    Right padding + Avg       &\underline{\textbf{69.0}} &\underline{\textbf{70.6}} &  \underline{\textbf{90.1}}  &   \underline{\textbf{64.4}}  &   \underline{\textbf{64.8}}   &  \underline{\textbf{69.6}}            &   \underline{\textbf{71.42}}              &     \textbf{71.6}       \\
    Left padding + Last       & \textbf{70.1}           &  \textbf{71.0}       &  90.1     &   \textbf{64.9}      &   \textbf{65.7}             &      \textbf{70.1}   &   \textbf{71.98}     &     67.9       \\
    Right padding + Last      & 67.4                    &  70.1                &  \textbf{90.2}      &   54.7               &   64.1                      &      56.1            &   67.10              &     \underline{\textbf{68.9}}            \\
    \hline
  \end{tabular}
  }
  \caption{The performance of code representations acquired in different ways on code search tasks. 
  Bold indicates optimal performance, and bold and underlined indicates suboptimal performance.
  }
  \label{embedding_type_res}
  % \vspace{-1.0em}
\end{table*}

To identify the most effective strategy for extracting code representations from decoder-only Transformer models, 
we investigated two approaches: decoder-only last token (\textbf{Last}) and decoder-only average (\textbf{Avg}). 
In addition, we examined how different padding strategies influence the quality and effectiveness of the resulting code representations.

Using CodeGen (350M) as an example, 
we conducted an ablation experiment on the downstream task of code search, 
with results presented in Table~\ref{embedding_type_res}. 
Our findings indicate that the optimal method for obtaining code representations varies with different padding strategies. 
When right padding is used, averaging all token embeddings yields better results than using only the embedding of the last token. 
Conversely, with left padding, the embedding of the last token outperforms the average of all embeddings. 
Considering results from both the CSN and CoSQA datasets, 
the decoder-only Transformer achieves the best representation with right padding and averaging all token embeddings. 
This setting was used in all other experiments.

\subsection{RQ3: Ablation study}

To investigate the effectiveness of our method, 
we designed ablation experiments to analyze the contributions of different components in CL4D. 

We trained three groups of models based on the CodeGen model:
1) the original Codegen;
2) the model trained using the contrastive learning method with in-batch negatives;
3) the model trained using the contrastive learning method with in-batch negatives and hard negatives.
We compared their zero-shot performance on downstream code understanding tasks, with results shown in Table~\ref{ablation_zero-shot}. 
The findings reveal that using in-batch negatives significantly enhances the code understanding ability of decoder-only models. 
Adding hard negatives further improves performance by approximately 1.5\%. 
Contrastive learning is expected to become an effective method for the decoder-only Transformer architecture to unify code understanding and code generation tasks.

\begin{table}[b]
  \centering
  \resizebox{0.95\columnwidth}{!}{
  \begin{tabular}{lccc}
    \hline
    \multirow{2}{*}{Method}         & \multicolumn{2}{c}{Code Search}      &  Clone Detection \\
    \cline{2-4}
                                             & CSN            &  CoSQA      &  POJ-104 \\
    \hline
    CL4D                                     & 72.00                    &  51.20                &  45.84           \\
    \quad - Hard Negative                          & 70.80                    &  50.40                &  44.65          \\
    \qquad- In-Batch Negative                      & 1.42                    &  0.45                &  13.20               \\
    \hline
  \end{tabular}
  }
  \caption{Ablation results of CL4D in the zero-shot setting.}
  \label{ablation_zero-shot}
\end{table}

\subsection{RQ4: Zero-shot}
We evaluated baseline models without fine-tuning on code search and clone detection tasks. 
As shown in Table~\ref{CL4D_res}, the zero-shot performance of encoder-only models on two downstream code understanding tasks improves with the introduction of more code structure-related tasks or scaling up strategies. 
Notably, UniXcoder and CodeSage leverage contrastive learning tasks during pre-training, 
significantly enhancing zero-shot performance compared to other encoder-only models. 
By contrast, decoder-only models underperform on downstream tasks related to code understanding, 
as their pre-training primarily focuses on next-token prediction rather than bidirectional semantic representation. 

The experimental results indicate that 
the extensive code knowledge acquired by pre-trained code understanding models remains underutilized. 
To leverage this knowledge for code understanding tasks, 
we applied the CL4D method described in Section~\ref{sec:method} for continued pre-training of these decoder-only models. 
The experimental outcomes, presented in Table~\ref{CL4D_res}, demonstrate that:
1) Our proposed method significantly enhances the zero-shot performance of decoder-only models on both code search and clone detection tasks, with average improvements ranging from 40\% to 60\%, and a maximum improvement of 75.90\%.
2) Our method enables the zero-shot performance of decoder-only models on various code understanding tasks to surpass that of encoder-only models of comparable size.
3) Our method enables decoder-only models to match the performance of same-sized, fine-tuned encoder-only state-of-the-art models on the CSN dataset, even without fine-tuning.

\begin{table*}[htbp]
  \centering
  \resizebox{\textwidth}{!}{
  \begin{tabular}{lllll}
    \hline
    \multirow{2}{*}{}            & \multirow{2}{*}{Method}                                    & \multicolumn{2}{c}{Code Search}       &  Clone Detection \\
    \cline{3-5}
                                 &                                                                     & CSN            &  CoSQA      &  POJ-104 \\
    \hline
    \multirow{4}*{Encoder-only}  & CodeBERT~\cite{DBLP:conf/emnlp/FengGTDFGS0LJZ20} (125M)             & 0.10                    &  0.24                &  20.38               \\
                                 & GraphCodeBERT~\cite{DBLP:conf/iclr/GuoRLFT0ZDSFTDC21} (125M)        & 11.26                   &  16.20               &  31.27               \\
                                 & UniXcoder~\cite{DBLP:conf/acl/GuoLDW0022} (125M)                    & 46.40                   &  42.11               &  42.08               \\
                                 & CodeSage~\cite{zhang2024codesage} (1.3B)                            & 71.24                   &  47.53               &  \textbf{73.07}               \\
    \hline
    \multirow{4}*{Decoder-only}  & CodeGPT~\cite{DBLP:conf/nips/LuGRHSBCDJTLZSZ21} (125M)              & 0.12                    &  0.04                &  9.41               \\
                                 & CodeGen~\cite{DBLP:conf/iclr/NijkampPHTWZSX23} (350M)               & 1.42                    &  0.45                &  13.20               \\
                                 & SantaCoder~\cite{DBLP:journals/corr/abs-2301-03988} (1.1B)          & 0.07                    &  0.11                &  15.57               \\
                                 & phi-1~\cite{DBLP:journals/corr/abs-2306-11644} (1.3B)               & 6.66                    &  6.75                &  27.62               \\
                                 & DeepSeek-Coder~\cite{DBLP:journals/corr/abs-2401-14196} (1.3B)                                       & 0.12                    &  0.63              &  16.51    \\
    \hdashline
    \multirow{4}*{Decoder-only}  & CodeGPT~\cite{DBLP:conf/nips/LuGRHSBCDJTLZSZ21} (125M) + CL4D       & 67.56 ($\uparrow$ 67.44)                   &  \textbf{53.49} ($\uparrow$ 53.45)               &  25.93 ($\uparrow$ 16.52)              \\
                                 & CodeGen~\cite{DBLP:conf/iclr/NijkampPHTWZSX23} (350M) + CL4D        & 71.97 ($\uparrow$ 70.55)                   &  51.18 ($\uparrow$ 50.73)               &  45.84 ($\uparrow$ 32.64)              \\
                                 & SantaCoder~\cite{DBLP:journals/corr/abs-2301-03988} (1.1B) + CL4D   & \underline{\textbf{74.18}} ($\uparrow$ 74.11)          &  52.82 ($\uparrow$ 52.71)               &  71.14 ($\uparrow$ 55.57)              \\
                                 & phi-1~\cite{DBLP:journals/corr/abs-2306-11644} (1.3B) + CL4D        & 73.21 ($\uparrow$ 66.55)  &  \underline{\textbf{52.95}} ($\uparrow$ 46.20)   &  69.99 ($\uparrow$ 42.37)              \\
                                 & DeepSeek-Coder~\cite{DBLP:journals/corr/abs-2401-14196} (1.3B) + CL4D                                &  \textbf{76.02} ($\uparrow$ 75.90)                        &   48.34 ($\uparrow$ 47.71)                                     &   \underline{\textbf{71.18}} ($\uparrow$ 54.67)   \\
    \hline
  \end{tabular}
  }
  \caption{Comparison of the performance of encoder-only models and decoder-only models on code understanding tasks in the zero-shot settings. 
  Bold indicates optimal performance, and bold and underlined indicates suboptimal performance.}
  \label{CL4D_res}
\end{table*}

\subsection{RQ5: Why does our approach work?}

\begin{figure*}[ht]
  \centering
  \includegraphics[width=0.9\textwidth]{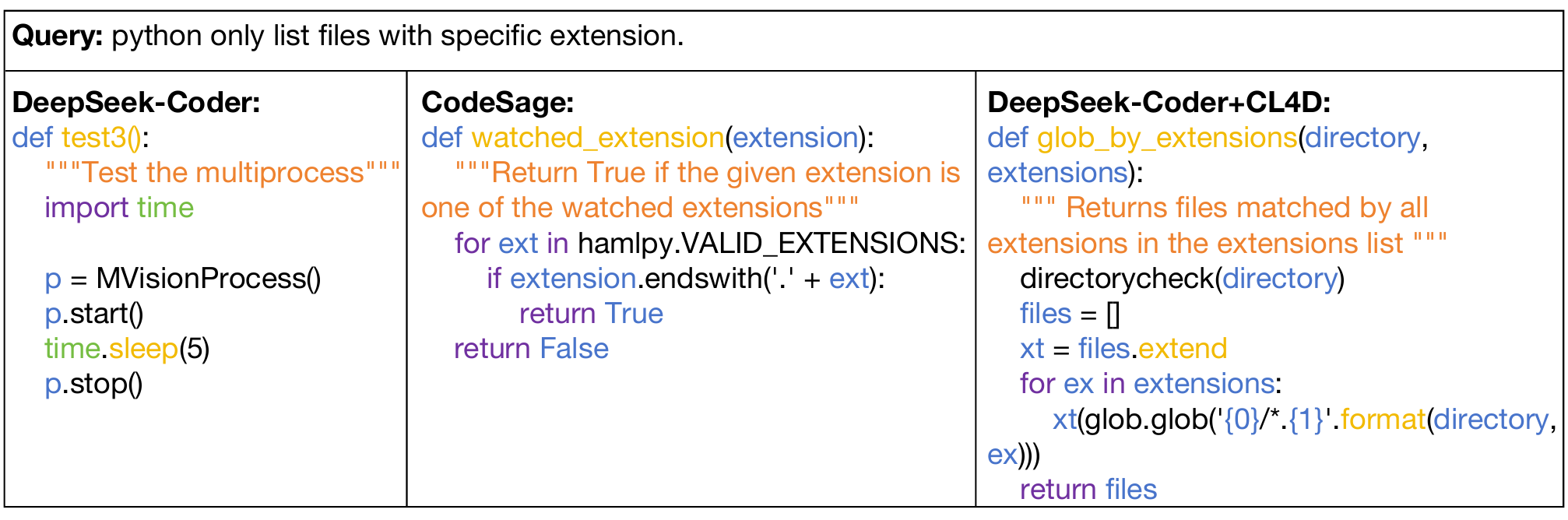}
  \caption{Case study on the code search task in the zero-shot setting.}
  \label{fig:case_study}
  \vspace{-1em}
\end{figure*}

\textbf{Case study}  
Figure~\ref{fig:case_study} displays the top-1 ranked code snippets returned by DeepSeek-Coder, CodeSage, and DeepSeek-Coder+CL4D for the query ``python only list files with specific extension''. 
DeepSeek-Coder+CL4D successfully returns the correct code snippet, which iterates through files and identifies those matching the specified extension. 
In contrast, the code snippets from DeepSeek-Coder and CodeSage are incorrect: DeepSeek-Coder returns a result entirely unrelated to the query, while CodeSage's snippet only identifies the extension without returning the corresponding files. 
These results indicate that continued pre-training of decoder-only models with CL4D significantly enhances their code understanding ability, allowing them to outperform encoder-only models of the same size.

\begin{figure*}[ht]
  \centering

  \subfigure[CodeGPT encodes CSN.]{
  \begin{minipage}{0.30\textwidth}
      \centering
      \includegraphics[width=\textwidth]{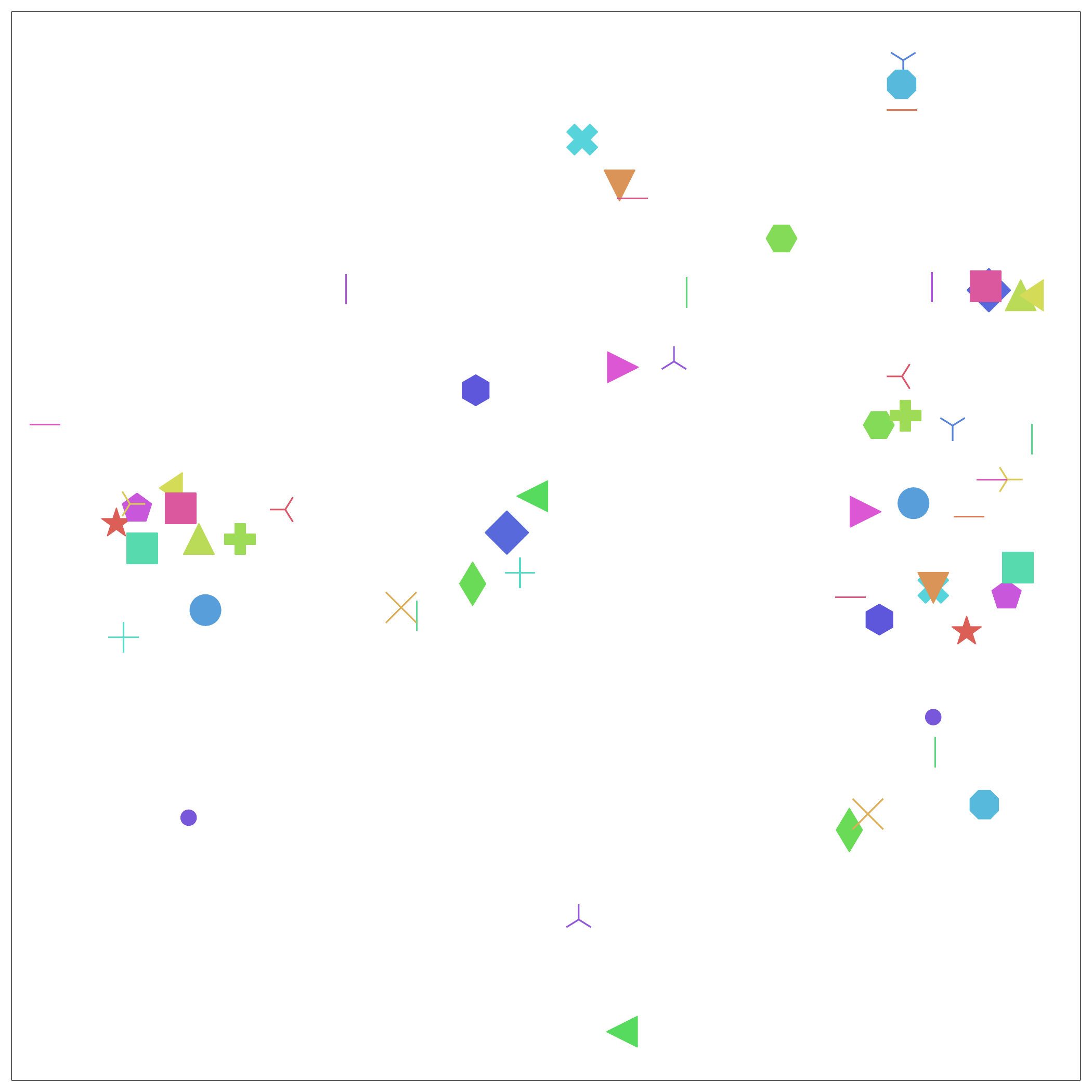}
  \end{minipage}
  }
  \subfigure[CodeGPT encodes CoSQA.]{
  \begin{minipage}{0.30\textwidth}
      \centering
      \includegraphics[width=\textwidth]{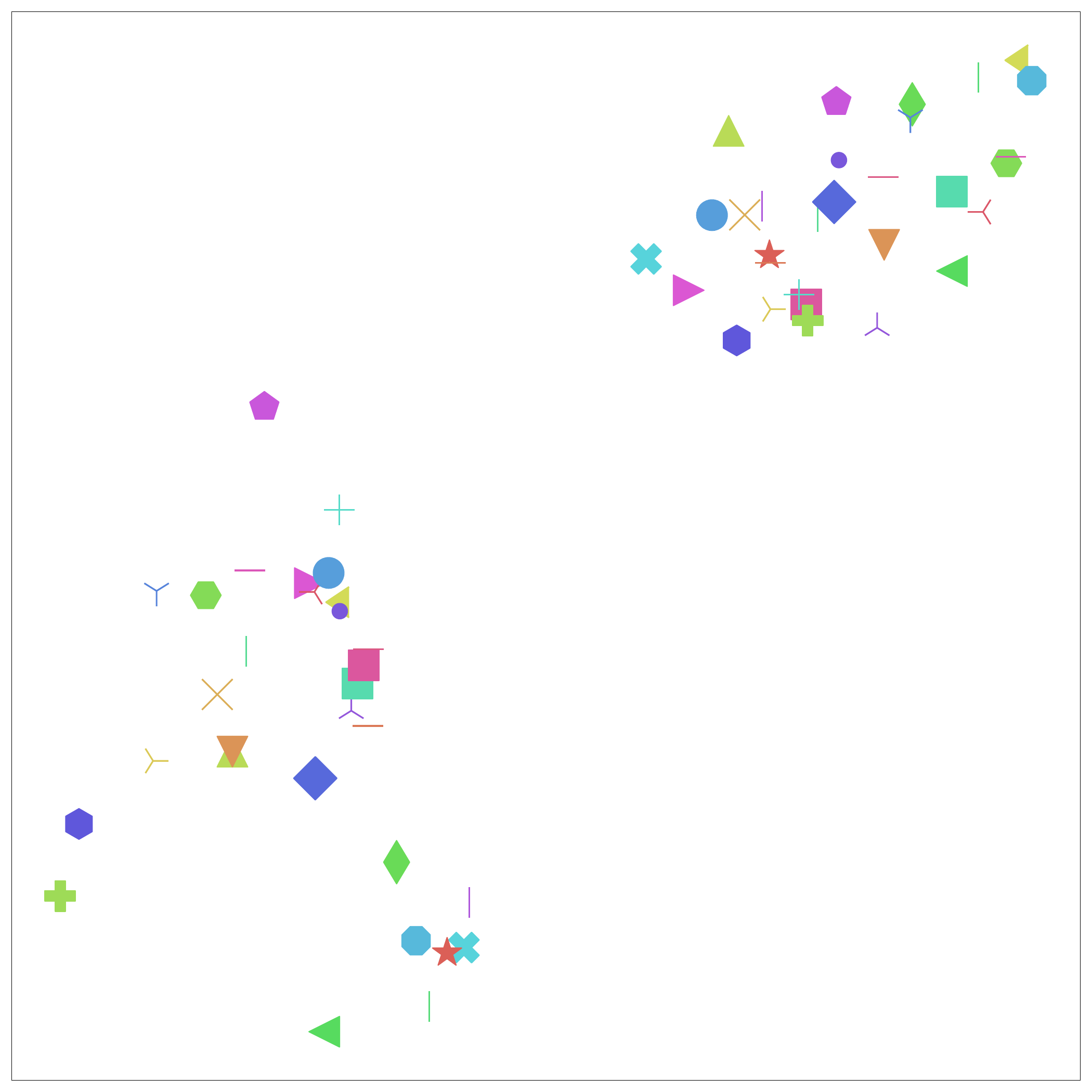}
  \end{minipage}
  }
  \subfigure[CodeGPT encodes POJ-104.]{
  \begin{minipage}{0.30\textwidth}
    \centering
    \includegraphics[width=\textwidth]{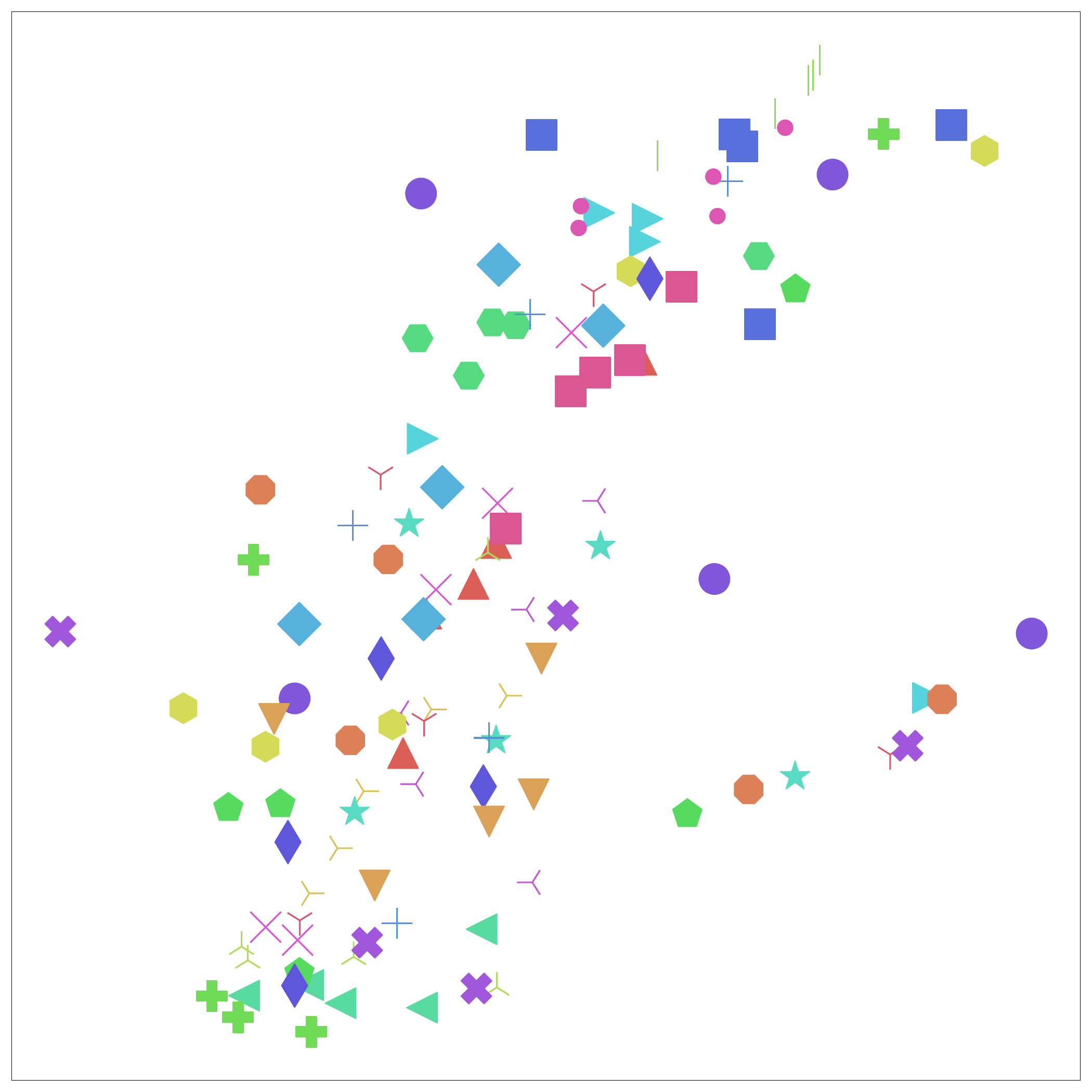}
  \end{minipage}
  }

  \vspace{0.3cm}

  \subfigure[CodeGPT+CL4D encodes CSN.]{
  \begin{minipage}{0.30\textwidth}
      \centering
      \includegraphics[width=\textwidth]{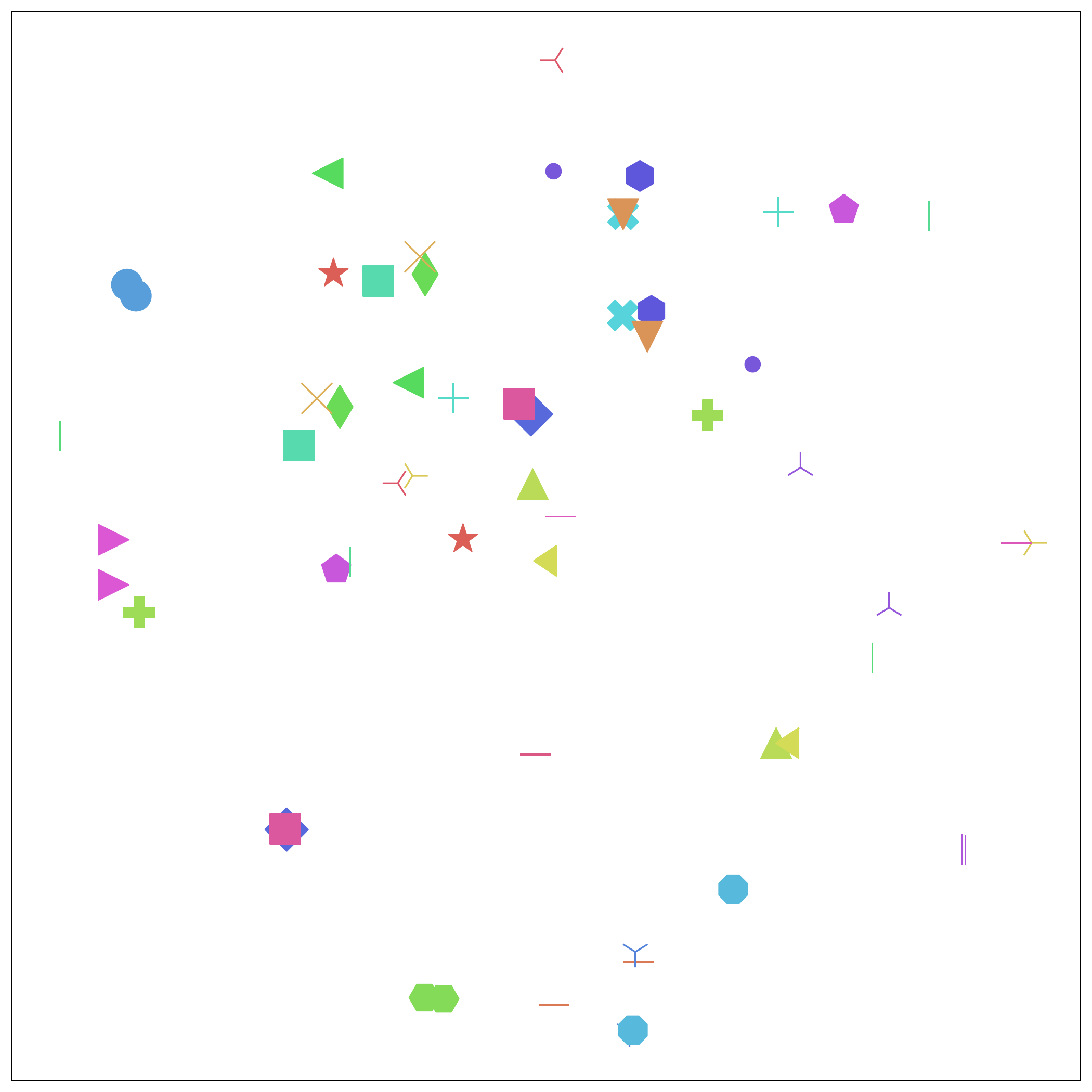}
  \end{minipage}
  }
  \subfigure[CodeGPT+CL4D encodes CoSQA.]{
  \begin{minipage}{0.30\textwidth}
      \centering
      \includegraphics[width=\textwidth]{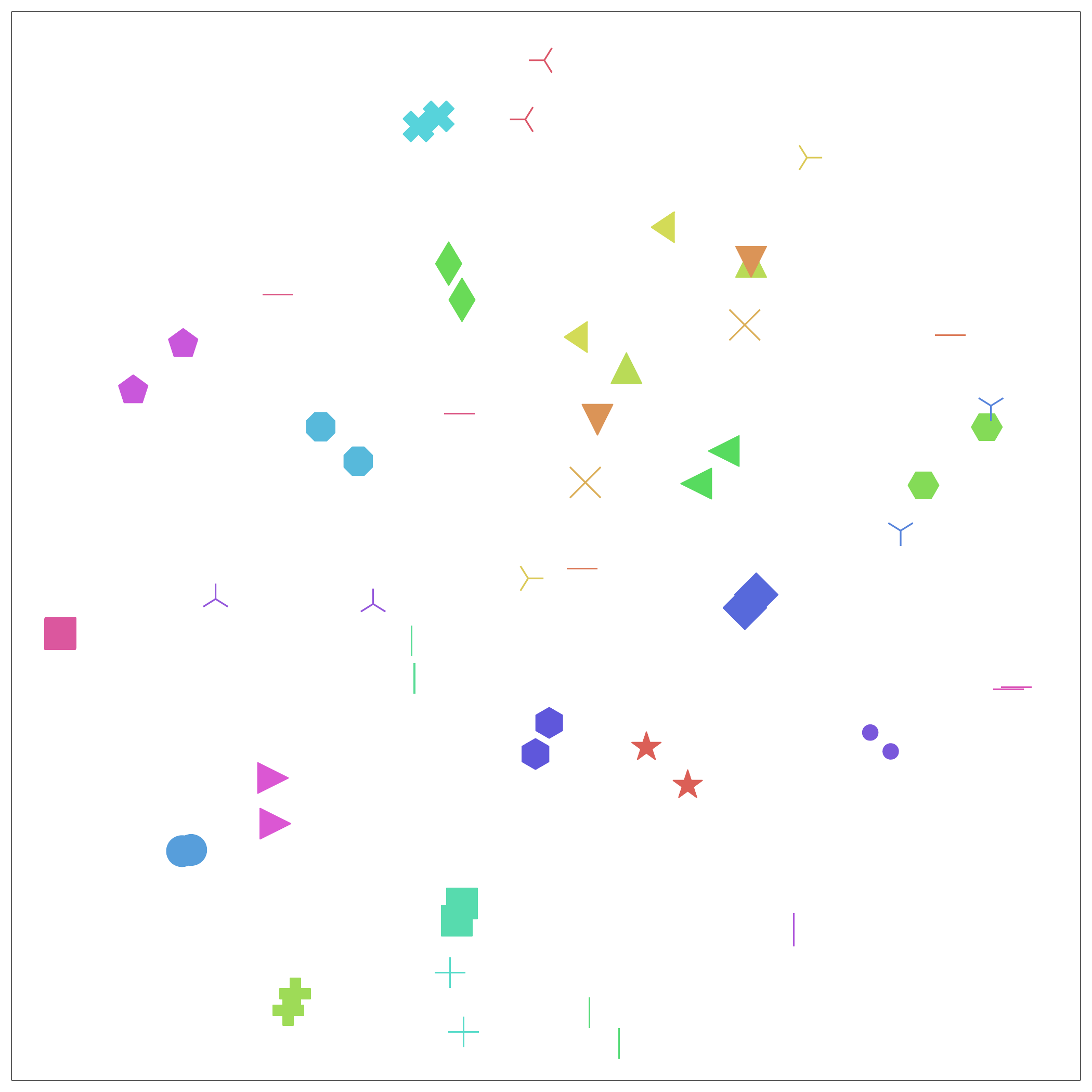}
  \end{minipage}
  }
  \subfigure[CodeGPT+CL4D encodes POJ-104.]{
  \begin{minipage}{0.30\textwidth}
    \centering
    \includegraphics[width=\textwidth]{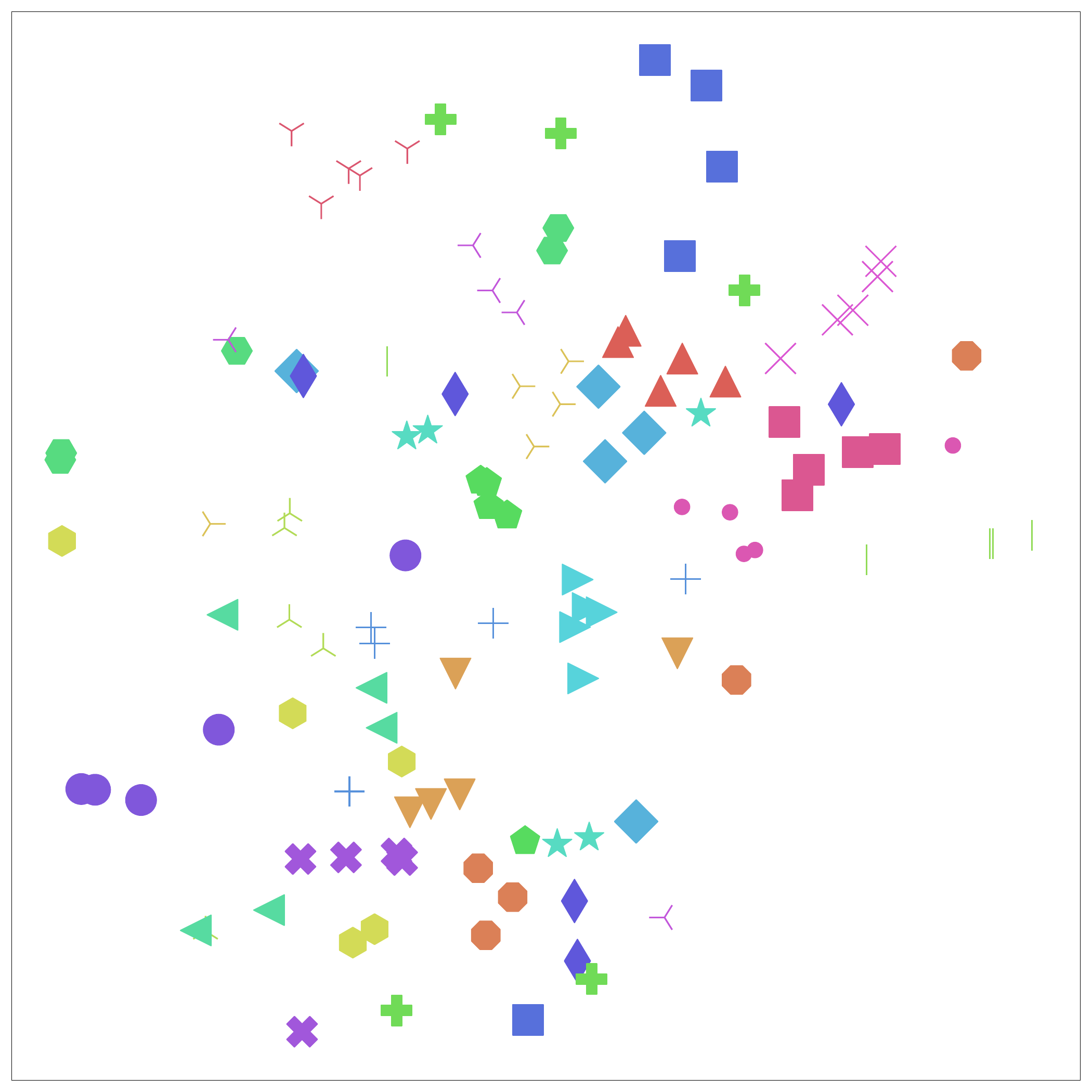}
  \end{minipage}
  }

  \caption{Visualization of code representations. 
  The first row shows the encoding results of CodeGPT, while the second row illustrates the encoding results of CodeGPT+CL4D. }
  \label{fig:visualization_res}
  \vspace{-1em}
\end{figure*}

\textbf{Visualization}
To investigate why CL4D effectively improves the performance of decoder-only models on code understanding tasks, 
we visualized the representation distribution of CodeGPT and CodeGPT+CL4D across the CSN, CoSQA, and POJ-104 datasets. 
For the CSN dataset, each query has only one correct answer, 
we randomly sampled 30 query-code pairs from the test set to form 30 categories, each containing 2 samples. 
The same approach was applied to the CoSQA dataset. 
For the POJ-104 test set, which has 24 problems, each with multiple correct answers, we sampled 5 correct answers per problem, forming 24 categories with 5 samples each. 

We then feed the sampled data into the models to obtain their high-dimensional feature space representations. 
Using t-SNE~\cite{van2008visualizing}, we reduce the dimensionality of these representations and visualize their distribution, as shown in Figure~\ref{fig:visualization_res}. 
Points with the same color and shape represent samples with the same semantics. 
Ideally, their representations in the vector space should be as close as possible. 
The results show that the representations encoded by the original decoder-only model are highly scattered, with semantically similar samples far apart in the vector space. 
The decoder-only model pre-trained with CL4D aligns the representations of semantically similar samples, bringing them closer together in the semantic space.

Overall, the visualization clearly demonstrates that our method significantly enhances the representation capability of the decoder-only models.

\section{Related works}
\textbf{Code Represention}
Recent work has focused on obtaining programming language representations for code understanding tasks. Models like CodeBERT~\cite{DBLP:conf/emnlp/FengGTDFGS0LJZ20}, StarEncoder~\cite{DBLP:journals/corr/abs-2305-06161}, CodeSage~\cite{zhang2024codesage}, and CoLSBERT~\cite{DBLP:journals/corr/abs-2402-12813} use the Masked Language Modeling (MLM) objective from NLP, applying it to serialized code. Recognizing the structural differences between code and text, some models enhance learning by incorporating code structures. For example, GraphCodeBERT~\cite{DBLP:conf/iclr/GuoRLFT0ZDSFTDC21} uses data flow, TreeBERT~\cite{DBLP:conf/uai/JiangZLLL21} utilizes the abstract syntax tree (AST), and SynCoBERT~\cite{wang2021syncobert} and UniXcoder~\cite{DBLP:conf/acl/GuoLDW0022} include serialized ASTs during training.

Although these models advance code understanding, they remain relatively small, typically trained on datasets with only millions of samples, and even the largest model, CoLSBERT, has only 1.5B parameters. While scaling up model size and training data can further improve performance, it also incurs substantial computational and resource costs. Inspired by the success of large-scale decoder-only code generation models, our work aims to leverage such pre-trained models to improve code understanding tasks, achieving state-of-the-art performance.

\textbf{Contrastive Learning for Code}
Contrastive learning has been widely employed in representation learning for deep learning networks. 
Recent studies have extended this paradigm to code corpora, 
proposing various methods to construct positive or negative samples to enhance code representation. 
For example, CoSQA~\cite{DBLP:conf/acl/HuangTSG0J0D20} constructs positive samples by rewriting the query, 
SynCoBERT~\cite{wang2021syncobert} and Code-MVP~\cite{DBLP:conf/naacl/WangWWWZLWL22} create positive sample pairs across different code modalities using diverse information obtained during program compilation, 
and UniXcoder~\cite{DBLP:conf/acl/GuoLDW0022}, inspired by SimCLR~\cite{DBLP:conf/icml/ChenK0H20}, uses dropout to construct positive samples in the latent vector space. 
Additionally, ChatDANCE~\cite{DBLP:conf/icsm/WangGSCCZWLZLZ23} uses large language models such as ChatGPT to synthesize positive code samples. 
CodeRetriever~\cite{DBLP:conf/emnlp/LiGSQZYQJCD22}, R2~\cite{DBLP:journals/corr/abs-2305-04508}, and CodeSage~\cite{zhang2024codesage} construct hard negatives for queries using rankers. 
In this paper, to mitigate the limited representation capability of decoder-only models by their unidirectional attention mechanism, 
we select code snippets with different semantics from the query but close in the representation space as hard negatives.

\section{Conclusion}
In this paper, we have presented a novel application of pre-trained decoder-only code generation models to code understanding tasks. 
We explore four distinct methods by which these decoder-only models develop code representations. 
Our study includes a comparative analysis of encoder-only and decoder-only models, examining their performance on downstream code understanding tasks in a zero-shot setting. 
To enhance the representational capabilities of the decoder-only model, 
we introduce a contrastive learning method, CL4D, which constructs in-batch negatives and hard negatives for training. 
Experimental results demonstrate that CL4D enables decoder-only models to significantly outperform encoder-only models in most tasks. 
Our findings indicate that decoder-only Transformer have the potential to unify code understanding and generation tasks.

\bibliography{aaai2026}

\end{document}